\documentclass[12pt]{revtex4}
\usepackage{amsmath,amssymb,bm}
\usepackage[toc,page]{appendix}
\usepackage{graphicx}
\usepackage{url}
\usepackage{hyperref,xcolor}

\DeclareGraphicsExtensions{.pdf,.png,.jpg,.eps}

\begin{document}
\title{The anomalous density of states and quasi-localized vibration through homogeneous thermalization of an inhomogeneous elastic system}
\author{Cunyuan Jiang$^{1,2,3}$}
\email{cunyuanjiang@sjtu.edu.cn}
\address{$^1$ School of Physics and Astronomy, Shanghai Jiao Tong University, Shanghai 200240, China}
\address{$^2$ Wilczek Quantum Center, School of Physics and Astronomy, Shanghai Jiao Tong University, Shanghai 200240, China}

\begin{abstract}
Amorphous solids are dynamically inhomogeneous due to in lack of translational symmetry and hence exhibit vibrational properties different from crystalline solids with anomalous low frequency vibrational density of states (VDOS) and related low temperature thermal properties. However, an interpretation of their origin from basic physical laws is still needed compared with rapidly progressed particle level investigations. In this work, we start with the quasi-equilibrium condition, which requires elastic potential energy to be homogeneously distributed even in an inhomogeneous elastic solid over long time observation. Analytical result shows that the anomalous low frequency VDOS behavior \(D(\omega) \propto \omega^4\) can be obtained when the quasi-equilibrium condition is satisfied on an inhomogeneous elastic system. Under high frequency after a crossover depending on the length scale of inhomogeneity, the power law of VDOS is changed to square \(D(\omega) \propto \omega^2\) which is Debye's law for crystalline solids. These features agree with recent particle level investigations. Our work suggest that the universal low frequency anomaly of amorphous solids can be considered as a result of homogeneous thermalization.
\end{abstract}

\maketitle
\section{Introduction}
Amorphous solids have been extensively observed with excess heat capacity and vibrational density of states (VDOS) than crystalline counterparts in the low temperature and frequency regime known as Boson peak (BP) since the anomaly was first reported in 1970s.\cite{PhysRevB.4.2029,Ahart2017,TsuneyoshiNakayama_2002} The anomaly of low frequency VDOS and low temperature thermal properties, which are also due to anomalous VDOS, are clearly because of inhomogeneity of amorphous solids. However, the mechanism that how inhomogeneity can leads to low frequency anomalous VDOS is still a controversial topic. 

During many decades of research, many theories have been proposed from various aspects to explain the question how inhomogeneity can leads to low frequency anomalous VDOS. For example, the inhomogeneous elasticity theory which used a field theoretical perturbative approach to consider the dressed phonon propagator in spatially fluctuating elastic medium.\cite{PhysRevLett.100.137402,PhysRevLett.98.025501} However, the inhomogeneous elasticity theory is considered to be underestimated to the microscopic mechanism.\cite{PhysRevLett.123.055501,PhysRevLett.130.236101} The quasi-localized vibration theory show that a vibrational instability of the spectrum of weakly interacting quasi-local harmonic modes creates the anomalous VDOS and predict \(D(\omega) \propto \omega^4\) at low frequency limit,\cite{PhysRevB.67.094203,PhysRevB.43.5039,PhysRevB.76.064206} and related sound damping effect.\cite{Baggioli_2022,Mizuno2025}. Even though the origin of quasi-local harmonic modes still need investigations. In addition, the self-consistent random matrix theory provide a microscopic field theoretical approach to grasp the low frequency \(\omega^4\) VDOS,\cite{PhysRevLett.130.236101} however the visualization of quasi-localized modes and their size effect are still needed to be discussed. The theory based on anharmonicity effect is also developed by considering Rayleigh damping of sound into field theoretical approach,\cite{PhysRevLett.122.145501} still the microscopic process is underestimated by the Rayleigh damping parameter. 

On the other side, the investigation based on experiments and simulations had also rapid progresses and provided more microscopic information about the origin of anomalous VDOS. Based on the experiments and simulations, the anomalous VDOS is confirmed to be contributed by transverse vibration.\cite{Ren2021,PhysRevLett.106.225501,Shintani2008} In addition, the low frequency power law \(D(\omega) \propto \omega^4\) is confirmed indicating the anomalous VDOS are quasi-localized as predicted by theories.\cite{10.21468/SciPostPhys.15.2.069,PhysRevLett.127.215504} By dispersing the anomalous low frequency vibrational mode onto particle level, simulation results revealed the vibrational anomaly is contributed by string-like dynamical defects which are a kind of soft spots with void space associated at the ends.\cite{Hu2022,PhysRevResearch.5.023055} From the real time simulation, the origin of anomalous VDOS can also be attributed by string-like motion which is a string of particles moving in same direction like a worm.\cite{PhysRevLett.80.2338,Zhang2011,10.1063/1.4769267} The relation between string-like dynamical defects and anomalous VDOS is also confirmed in experiment of active particles.\cite{PhysRevLett.133.188302} With the experiments and simulations results of string-like dynamical defects, the frequency of BP, where the power law start to change from \(\omega^4\) to \(\omega^2\), is found to be universally determined by the length of the string-like dynamical defects through a simple relation \(\omega_{BP} \sim c/\bar{l}\) with \(\bar{l}\) the average length and \(c\) the transverse speed of sound.\cite{PhysRevLett.133.188302,10.1063/5.0210057,Jiang_2024} 

The recent experiments and simulations progresses suggested that the low frequency anomalous VDOS is due to string-like quasi-localized dynamical soft spots, and the frequency of anomalous VDOS crossover is determined by their size. However, an interpretation from fundamental principle of how quasi-localized vibration and hence the anomalous VDOS can originate from dynamical soft spots is still needed. In this work, we study the relation between vibrational intensity and spatially fluctuating elasticity when quasi-equilibrium condition is satisfied, which requires the elastic potential energy density and hence the total energy density to be homogeneously distributed in space during long time observation. In real space, the analytic result show that vibrational intensity should be stronger at where elasticity is weaker to satisfy the quasi-equilibrium condition. In Fourier space, the analytic result of VDOS show that the power law change from \(\omega^4\) under low frequency to \(\omega^2\) under high frequency at the crossover depending on the length scale of fluctuation of elasticity through \(\omega_{BP} \sim \sqrt{3}c/(2 \bar{l})\). Our results provide an interpretation to the origin of low frequency anomalous VDOS in amorphous solids from fundamental quasi-equilibrium consideration, provide also an example connecting anomalous vibration phenomenon in inhomogeneous medium to the profound question of what will happen if an inhomogeneous medium is homogeneously thermalized.

\section{Analysis and discussions}
Let's consider a scaler field \(\phi(\boldsymbol{r},t)\) which denote the elastic vibrational displacement magnitude of a point \(\boldsymbol{r}\) in 3D space and at time \(t\). The full description of elastic displacement in 3D medium require vector field. However, because the components of different directions are relatively independently obeying elastic wave equation,\cite{thebook} one can use a scalar field for the convenience of analytic calculation. Elastic vibration is a result that the kinetic energy and potential energy convert to each other and hence elastic system have always \(E_{kin} \equiv E_{pot}\). For an elastic medium with fluctuating elasticity, the elastic potential energy density is defined through the spatial gradient of displacement magnitude and inhomogeneous elastic constant \( G(\boldsymbol{r}) \),
\begin{equation}
\mathcal{E}_\text{pot}(\boldsymbol{r}, t) = \frac{1}{2} G(\boldsymbol{r}) \left(\nabla \phi(\boldsymbol{r}, t)\right)^2.
\end{equation}
The quasi-equilibrium condition requires there is no energy current during long time observation, therefore the gradient of time averaged potential energy density should be zero everywhere, that is,
\begin{equation}
    \nabla \langle \mathcal{E}_\text{pot} \rangle_t (\boldsymbol{r}) \equiv 0, \label{mainequilibrium}
\end{equation}
where \(\langle \rangle_t\) denotes time average. The main line of this work is to obtain the quasi-localized vibration and anomalous VDOS from quasi-equilibrium condition Eq.\eqref{mainequilibrium} with inhomogeneous elasticity distribution \(G(\boldsymbol{r})\).

The scaler field \(\phi(\boldsymbol{r},t)\) can take a standard field description which is linear combination of eigenmodes with various wave vectors \(\boldsymbol{k}_n\) and their magnitude can be from zero to the maximum \(k_D\), the Debye's wave vector,
\begin{equation}
\phi(\boldsymbol{r}, t) = \sum_{\boldsymbol{k}_n} \left( \alpha_n(\boldsymbol{r}) e^{-i (\boldsymbol{k}_n \cdot \boldsymbol{r} - c |\boldsymbol{k}_n| t)} + \alpha_n^*(\boldsymbol{r}) e^{i (\boldsymbol{k}_n \cdot \boldsymbol{r} - c |\boldsymbol{k}_n| t)} \right). \label{mainfieldreal}
\end{equation}
Here \( \alpha_n (\boldsymbol{r}) \), a pure imaginary number, and its complex conjugate \( \alpha_n^* (\boldsymbol{r}) \) determine the vibration amplitude \(\alpha_n^* (\boldsymbol{r}) \alpha_n (\boldsymbol{r})\) at position \(\boldsymbol{r}\), \(\boldsymbol{k}_n = 2 \pi (n_1/L, n_2/L, n_3/L)\) are discrete wave vectors that satisfy the boundary condition with \(n_i\) integers and \(L\) the size of the system. In the language of field theory, \( \alpha_n (\boldsymbol{r}) \) and \( \alpha_n^* (\boldsymbol{r}) \) describe the annihilation and creation of a excitation with wave vector \(\boldsymbol{k}_n\) at position \(\boldsymbol{r}\), and \(\alpha_n^* (\boldsymbol{r}) \alpha_n (\boldsymbol{r})\) gives the number of excitation at position \(\boldsymbol{r}\). Here the speed of sound \(c\) is considered to be a constant for the plane waves with wave vector \(\boldsymbol{k}_n\). The vibration amplitude \( \alpha_n (\boldsymbol{r})\) here is allowed to change for different position \(\boldsymbol{r}\). With the expression of displacement magnitude \(\phi(\boldsymbol{r},t)\) in Eq.\eqref{mainfieldreal}, the time averaged potential density can be termed out to be, (See Sec.\ref{sec. Time averaged potential energy density} in SI.)
\begin{equation}
    \langle \mathcal{E}_\text{pot} \rangle_t (\boldsymbol{r}) = \pi G(\boldsymbol{r}) \sum_{\boldsymbol{k}_n} \left( \left( \nabla \alpha_n(\boldsymbol{r}) \right)^2 + |\boldsymbol{k}_n|^2\alpha_n(\boldsymbol{r})^2 \right). \label{mainpotdensityreal}
\end{equation}
Here \(\alpha_n (\boldsymbol{r})^2 \equiv \alpha_n (\boldsymbol{r}) \alpha_n^* (\boldsymbol{r})\) to make sure the energy is a positive number. By applying the quasi-equilibrium condition Eq.\eqref{mainequilibrium}, one can find the vibration amplitude \(\alpha_n (\boldsymbol{r})\) should satisfy the following relation with the fluctuating elastic constant,
\begin{equation}
G(\boldsymbol{r})^{-1} = \sum_{\boldsymbol{k}_n} \left( |\boldsymbol{k}_n|^2 \alpha_n(\boldsymbol{r})^2 +  \left(\nabla \alpha_n(\boldsymbol{r})\right)^2 \right). \label{mainrealspacerelation}
\end{equation}
Eq.\eqref{mainrealspacerelation} indicates that, where elasticity is weak, the vibration amplitude should be strong or the vibration amplitude should be strongly changed. In simulation works, the quasi-localized mode is defined as where vibration amplitude is strong in some areas compare with others although the local vibration amplitude were described by different ways, by particle level density of states in frequency domain,\cite{Hu2022,PhysRevResearch.5.023055,PhysRevLett.133.188302} or by string-like collective motion in time domain.\cite{PhysRevLett.80.2338,Zhang2011,10.1063/1.4769267} Therefore Eq.\eqref{mainrealspacerelation} explained the mechanism how soft spot can create quasi-localized mode, that quasi-localized mode appearing at soft spot is required by quasi-equilibrium condition Eq.\eqref{mainequilibrium}. If the vibration amplitude is same everywhere even elasticity is inhomogeneous, the time averaged elastic potential energy density and hence the total energy density would be also inhomogeneous, weaker at soft spot, which against the quasi-equilibrium condition Eq.\eqref{mainequilibrium}. When elasticity is homogeneous, \(G(\boldsymbol{r})^{-1} = const.\), Eq.\eqref{mainrealspacerelation} gives the vibrational intensity of each mode is also a constant as expected for plane waves in homogeneous mediums. 

From quasi-equilibrium condition Eq.\eqref{mainequilibrium}, real space relation between spatially fluctuating elasticity and vibrational intensity Eq.\eqref{mainrealspacerelation} can be derived which also dress the dynamical properties in Fourier space. To see that, we do the Fourier transform on both side of Eq.\eqref{mainrealspacerelation}, it becomes, (See Sec.\ref{sec. Fourier Transform of Elastic Constant Distribution Function} in SI.)
\begin{equation}
\tilde{G}^{-1}(\boldsymbol{q}) = \sum_{\boldsymbol{k}_n} |\boldsymbol{k}_n|^2 \tilde{\alpha}_n \star \tilde{\alpha}_n^* (\boldsymbol{q}) - [\boldsymbol{q} \tilde{\alpha}_n] \star [\boldsymbol{q} \tilde{\alpha}_n^*](\boldsymbol{q}), \label{mainkspacerelation}
\end{equation}
where \(\star\) denotes the convolution computation, and \(\tilde{\alpha}_n (\boldsymbol{q}) = \mathcal{F}[\alpha_n(\boldsymbol{r})]\) is the Fourier transform of vibrational intensity \(\alpha_n(\boldsymbol{r})\). In general, the Fourier transform of inverse elasticity, \(\tilde{G}^{-1}(\boldsymbol{q})\), can be wrote as a sum of Delta functions peaked at different wave vectors with different weight, \(\tilde{G}^{-1}(\boldsymbol{q}) = \sum_i p_i \delta(\boldsymbol{k}_{G,i} + \boldsymbol{q})\). In here, we use the simplest expression for avoiding tedious expansions,
\begin{equation}
    \tilde{G}^{-1}(\boldsymbol{q}) = \delta(\boldsymbol{k}_{G} + \boldsymbol{q}),
\end{equation}
which have fluctuating length scale \(1/k_G\) with \(k_G = |\boldsymbol{k}_G|\). With the given elasticity inverse distribution in wave vector space \(\tilde{G}^{-1}(\boldsymbol{q}) = \delta(\boldsymbol{k}_{G} + \boldsymbol{q})\), the Fourier transform of vibrational intensity function can be determined through Eq.\eqref{mainkspacerelation} to be a pure imaginary function, (See Sec.\ref{sec. Solve the vibrational amplitude through a given elastic constant distribution function} in SI.)
\begin{equation}
    \tilde{\alpha}_n(\boldsymbol{q}) = i \sqrt{\dfrac{1}{N}\dfrac{1}{|\boldsymbol{k}_n|^2 - \dfrac{3}{4}\boldsymbol{k}_G^2}} \delta(\dfrac{1}{2}\boldsymbol{k}_G + \boldsymbol{q}), \label{mainalphaq}
\end{equation}
with \(N\) is the normalization factor denoted the number of total modes.

The dynamical properties are usually described by current correlation function, which is defined as,\cite{ccf}
\begin{equation}
C(\boldsymbol{q}, t) = \partial_t \phi(\boldsymbol{q}, t) \partial_t \phi(-\boldsymbol{q}, 0),
\end{equation}
where \(\phi(\boldsymbol{q}, t) = \mathcal{F}_{\boldsymbol{r}}[\phi(\boldsymbol{r},t)]\) is the Fourier transform of the elastic vibrational displacement magnitude field. The spectra function \( C(\boldsymbol{q}, \omega) = \mathcal{F}_t [C(\boldsymbol{q},t)] \) can be obtained by applying a Fourier transform on time domain. According to the expression of \(\phi(\boldsymbol{r},t)\) in Eq.\eqref{mainfieldreal}, \( C(\boldsymbol{q}, \omega)\) can be obtained as, (See Sec.\ref{sec. Current Correlation Function} in SI.)
\begin{equation}
\begin{aligned}
C(\boldsymbol{q}, \omega) &= c^2 \sum_n \sum_m |\boldsymbol{k}_n| |\boldsymbol{k}_m| \tilde{\alpha}_n^*(\boldsymbol{k}_n - \boldsymbol{q}) \tilde{\alpha}_m(\boldsymbol{k}_m - \boldsymbol{q}) 2\pi \delta(\omega - c |\boldsymbol{k}_n|) \\
&\quad - |\boldsymbol{k}_n| |\boldsymbol{k}_m| \tilde{\alpha}_n^* (\boldsymbol{k}_n - \boldsymbol{q}) \tilde{\alpha}_m^* (\boldsymbol{k}_m + \boldsymbol{q}) 2\pi \delta(\omega - c |\boldsymbol{k}_n|) + c.c..\label{mainccffunction}
\end{aligned}
\end{equation}
where \(c.c.\) denotes the complex conjugate. Eq.\eqref{mainccffunction} contents four terms that corresponding to four momentum conserved dynamical processes. Using the annihilation and creation language in field theory, the first term describes that firstly annihilate and then create an excitation at \(\boldsymbol{q}\) in wave vector space, the second term describe creating two excitations at \(\boldsymbol{q}\) and \(-\boldsymbol{q}\), the third and forth are conjugate processes of the first two terms, that create and then annihilate an excitation at \(-\boldsymbol{q}\) and annihilate two excitations at \(\boldsymbol{q}\) and \(-\boldsymbol{q}\).

Then substitute \(\tilde{\alpha}_n(\boldsymbol{q})\) of Eq.\ref{mainalphaq} into Eq.\ref{mainccffunction}, one can obtain the current correlation function for a given fluctuating elasticity with fluctuating length scale \(1/k_G\), (See Sec.\ref{sec. Current correlation function for the given elastic constant distribution} in SI.)
\begin{equation}
\begin{aligned}
C(\boldsymbol{q}, \omega) &= \dfrac{c^2}{N} \sum_n \sum_m \dfrac{|\boldsymbol{k}_n| |\boldsymbol{k}_m|}{\sqrt{(|\boldsymbol{k}_n|^2 - \dfrac{3}{4}\boldsymbol{k}_G^2)(|\boldsymbol{k}_m|^2 - \dfrac{3}{4}\boldsymbol{k}_G^2)}} \\
&\quad \times \left[ \delta\left(\boldsymbol{k}_n - \left(\dfrac{1}{2}\boldsymbol{k}_G +\boldsymbol{q}\right)\right) \delta\left(\boldsymbol{k}_m - \left(\dfrac{1}{2}\boldsymbol{k}_G +\boldsymbol{q}\right)\right) 2\pi \delta(\omega - c |\boldsymbol{k}_n|) \right. \\
&\quad + \left. \delta\left(\boldsymbol{k}_n - \left(\dfrac{1}{2}\boldsymbol{k}_G +\boldsymbol{q}\right)\right) \delta\left(\boldsymbol{k}_m + \left(\dfrac{1}{2}\boldsymbol{k}_G +\boldsymbol{q}\right)\right) 2\pi \delta(\omega - c |\boldsymbol{k}_n|) \right] \\
& \quad + c.c., \label{mainccfgivenall}
\end{aligned}
\end{equation}
which is one of the main results of this work. It will be seen that Eq.\eqref{mainccfgivenall} can recover the Debye's law of VDOS \(D(\omega) \propto \omega^2 \) when \(k_G = 0\) elasticity is homogeneous and Eq.\eqref{mainccfgivenall} can gives the low frequency anomalous VDOS \(D(\omega) \propto \omega^4 \) when \(k_G \neq 0\) elasticity is inhomogeneous.

When \(k_G = 0\), the Delta functions require \(\boldsymbol{q} = \pm \boldsymbol{k}_n (\boldsymbol{k}_m)\) and \(|\boldsymbol{k}_n| \equiv |\boldsymbol{k}_m|\) for momentum conservation, Eq.\eqref{mainccfgivenall} becomes,
\begin{equation}
    C(\boldsymbol{q}, \omega) = \dfrac{2\pi c^2}{N}\delta(\omega - c|\boldsymbol{q}|).
\end{equation}
By doing the angular average and integral current correlation function over 3D wave vector space, the vibrational density of states (VDOS) can be obtained as,
\begin{equation}
    D(\omega) = \int dq \, \dfrac{2\pi c^2}{N}\delta(\omega - cq) 4\pi q^2 = \dfrac{8\pi^2}{N c} \omega^2.
\end{equation}

When \(\boldsymbol{k}_G \neq 0\), the momentum conservation still require \(|\boldsymbol{k}_n| \equiv |\boldsymbol{k}_m|\). Using the Delta function \(\delta(\omega - c|\boldsymbol{k}_n|)\), the factor in front is,
\begin{equation}
    \dfrac{|\boldsymbol{k}_n| |\boldsymbol{k}_m|}{\sqrt{(|\boldsymbol{k}_n|^2 - \dfrac{3}{4}\boldsymbol{k}_G^2)(|\boldsymbol{k}_m|^2 - \dfrac{3}{4}\boldsymbol{k}_G^2)}} = \dfrac{\omega^2}{\omega^2 - c^2 \dfrac{3}{4}k_G^2}.
\end{equation}
The Delta functions about wave vector give \(\boldsymbol{q} = \pm \boldsymbol{k}_n (\boldsymbol{k}_m) - 1/2 \boldsymbol{k}_G\) with a wave vector shift \(1/2 \boldsymbol{k}_G\) in stead of \(\boldsymbol{q} = \pm \boldsymbol{k}_n (\boldsymbol{k}_m)\) for homogeneous situation. This wave vector shift do not change frequency but only change the position of mode in wave vector space. If the wave vector shift is much smaller than Debye's wave vector, \(k_G \ll k_D\), then the effect can be negligible to the first order approximation when sum over all allowed wave vectors \(\sum_n \sum_m\). The error is just the mismatch of original Debye sphere, all allowed wave vectors smaller than \(k_D\), and shifted Debye sphere. The Debye's wave vector is in order of inter atomic distance inverse, and \(k_G\) is in order of length scale of soft spots inverse which in Experiments and simulation is about \(10\) atoms,\cite{Hu2022,PhysRevLett.133.188302,PhysRevResearch.5.023055} hence \(k_D \sim 10 k_G\). The approximation here weaken the shifting effect close to \(k_D\), which do not affect the presence of anomalous VDOS at low frequency and small wave vector. Therefore the current correlation function can be approximated to be, (See Sec.\ref{sec. Current correlation function for the given elastic constant distribution} in SI.)
\begin{equation}
    C(\boldsymbol{q}, \omega) \approx \dfrac{2 \pi c^2}{N} \dfrac{\omega^2}{\omega^2 - c^2 \dfrac{3}{4}k_G^2} \sum_n  \delta\left(\boldsymbol{k}_n - \boldsymbol{q}\right).
\end{equation}
Doing the angular average and integral current correlation function over 3D wave vector space, VDOS can be obtained as,
\begin{equation}
\begin{aligned}
    D(\omega) & \approx \dfrac{2 \pi c^2}{N} \int dq \, \dfrac{\omega^2}{\omega^2 + c^2 \dfrac{3}{4}k_G^2} 4\pi q^2 \\
    & \approx \dfrac{8\pi^2}{N c} \dfrac{\omega^4}{\omega^2 + c^2 \dfrac{3}{4}k_G^2}. \label{maindos}
\end{aligned}
\end{equation}

Under low frequency \(\omega \ll c\sqrt{\dfrac{3}{4}}k_G\),
\begin{equation}
    \dfrac{\omega^4}{\omega^2 + c^2 \dfrac{3}{4}k_G^2} \rightarrow \omega^4,
\end{equation}
and under high frequency \(\omega \gg c\sqrt{\dfrac{3}{4}}k_G\),
\begin{equation}
    \dfrac{\omega^4}{\omega^2 + c^2 \dfrac{3}{4}k_G^2} \rightarrow \omega^2.
\end{equation}
The frequency \(c\sqrt{\dfrac{3}{4}}k_G\) is where power law change, and hence the frequency of BP. About the \(\omega^4\) power law, it needs to be noticed that ordinary phonons have been subtracted here since the fluctuating elasticity \(\tilde{G}^{-1}(\boldsymbol{q}) = \delta(\boldsymbol{k}_G + \boldsymbol{q})\) has only one \(\delta\) function, which means it is only fluctuating part of elasticity.

Another way to see the effect of nonzero \(k_G\) is to limit that there is only one allowed mode in the displacement intensity field \(\phi(\boldsymbol{r}, t)\), that means,
\begin{equation}
\phi(\boldsymbol{r}, t) = \alpha_0(\boldsymbol{r}) e^{-i (\boldsymbol{k}_0 \cdot \boldsymbol{r} - c |\boldsymbol{k}_0| t)} + \alpha_0^*(\boldsymbol{r}) e^{i (\boldsymbol{k}_0 \cdot \boldsymbol{r} - c |\boldsymbol{k}_0| t)}.
\end{equation}
Then using Eq.\eqref{mainccfgivenall} the current correlation function becomes,
\begin{equation}
    C(\boldsymbol{q}, \omega) = \dfrac{2\pi c^2}{N} \dfrac{k_0^2}{k_0^2 + \dfrac{3}{4}k_G^2} \delta\left(\boldsymbol{k}_0 - \left(\dfrac{1}{2}\boldsymbol{k}_G + \boldsymbol{q}\right)\right) \delta(\omega - ck_0).
\end{equation}
Therefore the \(\boldsymbol{k}_G\) dressed vibrational mode will appear only at,
\begin{equation}
    C\left(\boldsymbol{k}_0 - \dfrac{1}{2}\boldsymbol{k}_G, ck_0 \right) = \dfrac{2\pi c^2}{N} \dfrac{k_0^2}{k_0^2 + \dfrac{3}{4}k_G^2},
\end{equation}
instead of standard acoustic phonon mode \(\left(\boldsymbol{k}_0, ck_0 \right)\). The angle between \(\boldsymbol{k}_G\) and \(\boldsymbol{k}_0\) can be any number in \([0,\pi]\). Therefore in the scalar wave vector plot of current correlation function \(C(q,\omega)\) which is usually used for the presentation of experiments and simulations results,\cite{Hu2022,PhysRevLett.133.188302,PhysRevResearch.5.023055} the signal of \(C(q,\omega)\) can cover a region \([|\dfrac{1}{2}k_G - k_0|, |\dfrac{1}{2}k_G + k_0|]\) in wave vector axis but only a point \(ck_0\) in frequency axis. This current correlation function smearing effect on wave vector axis can be considered to interpret the origin of quasi-flat dispersion relation observed in experiments and simulations results.\cite{Hu2022,PhysRevLett.133.188302,PhysRevResearch.5.023055}

As mentioned in the beginning, the frequency of BP which the anomalous VDOS \(\omega^4\) converted to \(\omega^2\) can be estimated from the length scale of string-like dynamical defects or string-like motion, by \(\omega_{BP} \sim c/\bar{l}\). The result of Eq.\eqref{maindos} is in good agreement with the phenomenological model.\cite{PhysRevLett.133.188302,10.1063/5.0210057,Jiang_2024} Here the expression of DOS in Eq.\eqref{maindos} is derived from quasi-equilibrium condition in Eq.\eqref{mainequilibrium} without phenomenological input. The term \(c^2 \dfrac{3}{4}k_G^2\) in denominator of Eq.\eqref{maindos} is vital to generate \(\omega^4\) low frequency anomalous VDOS, and it can be found the term which come from the gradient term \((\nabla \alpha_n(\boldsymbol{r}))^2\) in the time averaged elastic potential energy density in Eq.\eqref{mainpotdensityreal}. The term coming from time averaged elastic potential energy density indicates that the low frequency anomalous VDOS is directly a result of inhomogeneous vibrational intensity. The inhomogeneous vibrational intensity is naturally a result of the quasi-equilibrium condition which require elastic potential energy density and hence total energy density to be homogeneously distributed in even an inhomogeneous elastic system. This provide the origin of low frequency anomalous VDOS in amorphous solids a profound interpretation from quasi-equilibrium consideration.

\section{Summary}
In this work, the low frequency anomalous VDOS well observed in amorphous solids are explained from the consideration of homogeneous energy density distribution even the elasticity of medium is inhomogeneous. The results suggest that the quasi-localized vibration should appear at soft spots for keeping energy density distribution to be homogeneous during long time of observation. As a result of presence of quasi-localized vibration, the anomalous VDOS \(D(\omega) \propto \omega^4\) should appear at low frequency region lower than a crossover frequency \(c\sqrt{3/4} k_G\) determined by the typical length scale of soft spots \(1/k_G\). The results also suggest that presence of quasi-localized vibration can smear the wave vector of ordinary linear dispersion relation which may be used for understanding the origin of quasi-flat band observed recently in experiments and simulations. Our work explained the origin of quasi-localized vibration and low frequency anomalous VDOS in amorphous solids from quasi-equilibrium consideration, explained its size size dependent relation and possible effect on dispersion relation. Our work suggest that low frequency dynamic anomaly in amorphous solids can be considered as a result of homogeneously thermalization on inhomogeneous medium.

\section*{Acknowledgments}
The author would like to thank Matteo Baggioli for very illuminating discussions. The author acknowledge the support of the Shanghai Municipal Science and Technology Major Project (Grant No. 2019SHZDZX01).

\clearpage
\bibliography{main}

\newpage
\appendix 
\clearpage
\section*{\Large Supplementary Information}

\section{Time averaged potential energy density}\label{sec. Time averaged potential energy density}

Given the scalar field \( \phi(\boldsymbol{r}, t) \) which denotes displacement magnitude in an inhomogeneous elastic medium,
\begin{equation}
\phi(\boldsymbol{r}, t) = \sum_{\boldsymbol{k}_n} \left( \alpha_n(\boldsymbol{r}) e^{-i (\boldsymbol{k}_n \cdot \boldsymbol{r} - c |\boldsymbol{k}_n| t)} + \alpha_n^*(\boldsymbol{r}) e^{i (\boldsymbol{k}_n \cdot \boldsymbol{r} - c |\boldsymbol{k}_n| t)} \right),
\end{equation}
where \( \alpha_n (\boldsymbol{r}) \) is a imaginary number and its complex conjugate \( \alpha_n^* (\boldsymbol{r}) \) determine the vibration amplitude \( \sqrt{\alpha_n (\boldsymbol{r}) \alpha_n^* (\boldsymbol{r})} \) at position \(\boldsymbol{r}\), \(\boldsymbol{k}_n = 2 \pi (n_1/L, n_2/L, n_3/L)\) are discrete wave vectors that satisfy the boundary condition with \(n_i\) integers and \(L\) the size of the system. In the language of field theory, \( \alpha_n (\boldsymbol{r}) \) and (\( \alpha_n^* (\boldsymbol{r}) \)) served as annihilation (creation) operator which annihilate (create) an excitation at position \(\boldsymbol{r}\). Here the speed of sound \(c\) is considered to be a constant for the plane waves with wave vector \(\boldsymbol{k}_n\).

The potential energy density is defined through the spatial gradient of displacement magnitude and inhomogeneous elastic constant \( G(\boldsymbol{r}) \),
\begin{equation}
\mathcal{E}_\text{pot}(\boldsymbol{r}, t) = \frac{1}{2} G(\boldsymbol{r}) \left(\nabla \phi(\boldsymbol{r}, t)\right)^2.
\end{equation}
The spatial gradient of \( \phi \) is,
\begin{equation}
\begin{aligned}
\nabla \phi(\boldsymbol{r}, t) &= \sum_{\boldsymbol{k}_n} \left( \nabla \alpha_n(\boldsymbol{r}) e^{-i (\boldsymbol{k}_n \cdot \boldsymbol{r} - c |\boldsymbol{k}_n| t)} + \alpha_n(\boldsymbol{r}) \left( -i \boldsymbol{k}_n e^{-i (\boldsymbol{k}_n \cdot \boldsymbol{r} - c |\boldsymbol{k}_n| t)} \right) \right. \\
&\quad + \left. \nabla \alpha_n^*(\boldsymbol{r}) e^{i (\boldsymbol{k}_n \cdot \boldsymbol{r} - c |\boldsymbol{k}_n| t)} + \alpha_n^*(\boldsymbol{r}) \left( i \boldsymbol{k}_n e^{i (\boldsymbol{k}_n \cdot \boldsymbol{r} - c |\boldsymbol{k}_n| t)} \right) \right).
\end{aligned}
\end{equation}
Combine the terms with the same exponential,
\begin{equation}
\begin{aligned}
\nabla \phi(\boldsymbol{r}, t) &= \sum_{\boldsymbol{k}_n} \left( \left( \nabla \alpha_n(\boldsymbol{r}) - i \alpha_n(\boldsymbol{r}) \boldsymbol{k}_n \right) e^{-i (\boldsymbol{k}_n \cdot \boldsymbol{r} - c |\boldsymbol{k}_n| t)} \right. \\
&\quad + \left. \left( \nabla \alpha_n^*(\boldsymbol{r}) + i \alpha_n^*(\boldsymbol{r}) \boldsymbol{k}_n \right) e^{i (\boldsymbol{k}_n \cdot \boldsymbol{r} - c |\boldsymbol{k}_n| t)} \right) \\
&= \sum_{\boldsymbol{k}_n} 2 \text{Re} \left[ \left( \nabla \alpha_n(\boldsymbol{r}) - i \alpha_n(\boldsymbol{r}) \boldsymbol{k}_n \right) e^{-i (\boldsymbol{k}_n \cdot \boldsymbol{r} - c |\boldsymbol{k}_n| t)} \right],
\end{aligned}
\end{equation}
where \(\text{Re}[]\) denotes taking the real part due to complex conjugate properties. Let's take \(c_n = \left( \nabla \alpha_n(\boldsymbol{r}) - i \alpha_n(\boldsymbol{r}) \boldsymbol{k}_n \right)\) and \( \theta_n = (\boldsymbol{k}_n \cdot \boldsymbol{r} - c |\boldsymbol{k}_n| t) \) for convenience. Using Euler's equation, the gradient can be rearranged in a compact form,
\begin{equation}
    \nabla \phi(\boldsymbol{r}, t) = 2 \sum_{\boldsymbol{k}_n} (\text{Re}[c_n] \cos\theta_n - \text{Im}[c_n] \sin\theta_n).
\end{equation}
The potential energy density is therefore,
\begin{equation}
\begin{aligned}
\mathcal{E}_\text{pot}(\boldsymbol{r}, t) &= \frac{1}{2} G(\boldsymbol{r}) \left(\nabla \phi(\boldsymbol{r}, t)\right)^2 \\
&= \frac{1}{2} G(\boldsymbol{r}) \left( 2 \sum_{\boldsymbol{k}_n} (\text{Re}[c_n] \cos\theta_n - \text{Im}[c_n] \sin\theta_n) \right) \\
&\quad \times \left( 2 \sum_{\boldsymbol{k}_m} (\text{Re}[c_m] \cos\theta_m - \text{Im}[c_m] \sin\theta_m) \right) \\
&= 2 G(\boldsymbol{r}) \sum_{\boldsymbol{k}_n} \sum_{\boldsymbol{k}_m} \left( \text{Re}[c_n] \cos\theta_n \text{Re}[c_m] \cos\theta_m + 
\text{Im}[c_n] \sin\theta_n \text{Im}[c_m] \sin\theta_m \right. \\
&\quad - \text{Im}[c_n] \sin\theta_n \text{Re}[c_m] \cos\theta_m - \text{Im}[c_m] \sin\theta_m \text{Re}[c_n] \cos\theta_n \left.\right).
\end{aligned}
\end{equation}

The time-averaged potential energy density can be obtained by integrating over time \(t \),
\begin{equation}
\begin{aligned}
    \langle \mathcal{E}_\text{pot}(\boldsymbol{r}) \rangle_t &= \dfrac{1}{T} \int_0^T dt \mathcal{E}_\text{pot}(\boldsymbol{r}, t) \\
    & = 2 G(\boldsymbol{r}) \sum_{\boldsymbol{k}_n} \sum_{\boldsymbol{k}_m} \left( \int_0^T dt \text{Re}[c_n] \cos\theta_n \text{Re}[c_m] \cos\theta_m + 
    \int_0^T dt \text{Im}[c_n] \sin\theta_n \text{Im}[c_m] \sin\theta_m \right. \\
    &\quad - \int_0^T dt \text{Im}[c_n] \sin\theta_n \text{Re}[c_m] \cos\theta_m - \int_0^T dt \text{Im}[c_m] \sin\theta_m \text{Re}[c_n] \cos\theta_n \left.\right) \\
    & = 2 G(\boldsymbol{r}) \sum_{\boldsymbol{k}_n} \left( \text{Re}[c_n]^2 \int_0^T dt \cos^2\theta_n + \text{Im}[c_n]^2 \int_0^T dt \sin^2\theta_n \right) \\
    & = G(\boldsymbol{r}) \sum_{\boldsymbol{k}_n} \left( \text{Re}[c_n]^2 + \text{Im}[c_n]^2 \right).
\end{aligned}
\end{equation}
Here the orthogonality of trigonometric functions is used. They are,
\begin{equation}
\begin{aligned}
    & \int_0^T dt \sin\theta_n \sin\theta_m = \delta_{n,m} \dfrac{\pi}{2}, \\
    & \int_0^T dt \cos\theta_n \cos\theta_m = \delta_{n,m} \dfrac{\pi}{2}, \\
    & \int_0^T dt \cos\theta_n \sin\theta_m = 0.
\end{aligned}
\end{equation}
The vibrational amplitude \(\alpha_n (\boldsymbol{r})\) can be chosen to be a peal imaginary number, therefore,
\begin{equation}
    \langle \mathcal{E}_\text{pot}(\boldsymbol{r}) \rangle_t = \pi G(\boldsymbol{r}) \sum_{\boldsymbol{k}_n} \left( \left( \nabla \alpha_n(\boldsymbol{r}) \right)^2 + |\boldsymbol{k}_n|^2\alpha_n(\boldsymbol{r})^2 \right). \label{potdensityfunction}
\end{equation}
Note that \(\left( \nabla \alpha_n(\boldsymbol{r}) \right)^2 \equiv \nabla \alpha_n(\boldsymbol{r})\cdot \nabla \alpha_n^*(\boldsymbol{r})\) and \(\alpha_n(\boldsymbol{r}) ^2 \equiv \alpha_n(\boldsymbol{r}) \alpha_n^*(\boldsymbol{r})\).

\section{Current Correlation Function} \label{sec. Current Correlation Function}
Dynamical properties of the system can be obtained through a current correlation function, in time domain, it is defined as,
\begin{equation}
C(\boldsymbol{q}, t) = \partial_t \phi(\boldsymbol{q}, t) \partial_t \phi(-\boldsymbol{q}, 0).\label{ccf}
\end{equation}
Compute the time derivative of \( \phi \),
\begin{equation}
\partial_t \phi(\boldsymbol{r}, t) = \sum_n \left( -i c |\boldsymbol{k}_n| \alpha_n(\boldsymbol{r}) e^{-i (\boldsymbol{k}_n \cdot \boldsymbol{r} - c |\boldsymbol{k}_n| t)} + i c |\boldsymbol{k}_n| \alpha_n^*(\boldsymbol{r}) e^{i (\boldsymbol{k}_n \cdot \boldsymbol{r} - c |\boldsymbol{k}_n| t)} \right).
\end{equation}
The Fourier transform on the spatial domain is given by,
\begin{equation}
\partial_t \phi(\boldsymbol{q}, t) = \int d^3r \, \partial_t \phi(\boldsymbol{r}, t) e^{-i \boldsymbol{q} \cdot \boldsymbol{r}}.
\end{equation}
Substitute the expression for \(\partial_t \phi(\boldsymbol{r}, t)\),
\begin{equation}
\begin{aligned}
\partial_t \phi(\boldsymbol{q}, t) &= \int d^3r \sum_n \left( -i c |\boldsymbol{k}_n| \alpha_n(\boldsymbol{r}) e^{-i (\boldsymbol{k}_n \cdot \boldsymbol{r} - c |\boldsymbol{k}_n| t)} + i c |\boldsymbol{k}_n| \alpha_n^*(\boldsymbol{r}) e^{i (\boldsymbol{k}_n \cdot \boldsymbol{r} - c |\boldsymbol{k}_n| t)} \right) \\
&\quad \times e^{-i \boldsymbol{q} \cdot \boldsymbol{r}}
\end{aligned}
\end{equation}
Separate the integrals,
\begin{equation}
\begin{aligned}
\partial_t \phi(\boldsymbol{q}, t) &= \sum_n \left( \int d^3r \, (-i c |\boldsymbol{k}_n| \alpha_n(\boldsymbol{r}) e^{-i ((\boldsymbol{k}_n + \boldsymbol{q}) \cdot \boldsymbol{r} - c |\boldsymbol{k}_n| t)}) \right. \\
&\quad + \left. \int d^3r \, (i c |\boldsymbol{k}_n| \alpha_n^*(\boldsymbol{r}) e^{-i ((\boldsymbol{k}_n - \boldsymbol{q}) \cdot \boldsymbol{r} + c |\boldsymbol{k}_n| t)}) \right).
\end{aligned}
\end{equation}
The Fourier transform of \(\alpha_n(\boldsymbol{r})\) and \(\alpha_n^*(\boldsymbol{r})\) are,
\begin{equation}
\begin{aligned}
    & \tilde{\alpha}_n(\boldsymbol{q}) = \int d^3r \alpha_n(\boldsymbol{r}) e^{-i \boldsymbol{q} \cdot \boldsymbol{r}} \\
    & \tilde{\alpha}_n^* (\boldsymbol{q}) = \int d^3r \alpha_n^*(\boldsymbol{r}) e^{-i \boldsymbol{q} \cdot \boldsymbol{r}}.
\end{aligned}
\end{equation}
In field theory, they are annihilation and creation operator on momentum space. By replacing \(\boldsymbol{q}\) to \(\boldsymbol{k}_n \pm \boldsymbol{q}\) in the above expression, we have,
\begin{equation}
\partial_t \phi(\boldsymbol{q}, t) = \sum_n \left( -i c |\boldsymbol{k}_n| \tilde{\alpha}_n(\boldsymbol{k}_n + \boldsymbol{q}) e^{-i c |\boldsymbol{k}_n| t} + i c |\boldsymbol{k}_n| \tilde{\alpha}_n^*(\boldsymbol{k}_n - \boldsymbol{q}) e^{i c |\boldsymbol{k}_n| t} \right),
\end{equation}
Now, compute \(\partial_t \phi(-\boldsymbol{q}, 0)\),
\begin{equation}
\partial_t \phi(-\boldsymbol{q}, 0) = \sum_n \left( -i c |\boldsymbol{k}_n| \tilde{\alpha}_n(\boldsymbol{k}_n - \boldsymbol{q}) + i c |\boldsymbol{k}_n| \tilde{\alpha}_n^*(\boldsymbol{k}_n + \boldsymbol{q}) \right).
\end{equation}

The current correlation function in time domain is therefore \( C(\boldsymbol{q}, t) \),
\begin{equation}
\begin{aligned}
C(\boldsymbol{q}, t) &= -\sum_n \left( -i c |\boldsymbol{k}_n| \tilde{\alpha}_n(\boldsymbol{k}_n + \boldsymbol{q}) e^{-i c |\boldsymbol{k}_n| t} + i c |\boldsymbol{k}_n| \tilde{\alpha}_n^*(\boldsymbol{k}_n - \boldsymbol{q}) e^{i c |\boldsymbol{k}_n| t} \right) \\
&\quad \times \sum_m \left( -i c |\boldsymbol{k}_m| \tilde{\alpha}_m(\boldsymbol{k}_m - \boldsymbol{q}) + i c |\boldsymbol{k}_m| \tilde{\alpha}_m^*(\boldsymbol{k}_m + \boldsymbol{q}) \right).
\end{aligned}
\end{equation}
Expand the product,
\begin{equation}
\begin{aligned}
C(\boldsymbol{q}, t) &= \sum_n \sum_m \left( (-i c |\boldsymbol{k}_n| \tilde{\alpha}_n(\boldsymbol{k}_n + \boldsymbol{q}) e^{-i c |\boldsymbol{k}_n| t})(-i c |\boldsymbol{k}_m| \tilde{\alpha}_m(\boldsymbol{k}_m - \boldsymbol{q})) \right. \\
&\quad + (-i c |\boldsymbol{k}_n| \tilde{\alpha}_n(\boldsymbol{k}_n + \boldsymbol{q}) e^{-i c |\boldsymbol{k}_n| t})(i c |\boldsymbol{k}_m| \tilde{\alpha}_m^*(\boldsymbol{k}_m + \boldsymbol{q})) \\
&\quad + (i c |\boldsymbol{k}_n| \tilde{\alpha}_n^*(\boldsymbol{k}_n - \boldsymbol{q}) e^{i c |\boldsymbol{k}_n| t})(-i c |\boldsymbol{k}_m| \tilde{\alpha}_m(\boldsymbol{k}_m - \boldsymbol{q})) \\
&\quad + \left. (i c |\boldsymbol{k}_n| \tilde{\alpha}_n^*(\boldsymbol{k}_n - \boldsymbol{q}) e^{i c |\boldsymbol{k}_n| t})(i c |\boldsymbol{k}_m| \tilde{\alpha}_m^*(\boldsymbol{k}_m + \boldsymbol{q})) \right).
\end{aligned}
\end{equation}
Simplify the expression,
\begin{equation}
\begin{aligned}
C(\boldsymbol{q}, t) &= c^2 \sum_n \sum_m \left( - |\boldsymbol{k}_n| |\boldsymbol{k}_m| \tilde{\alpha}_n(\boldsymbol{k}_n + \boldsymbol{q}) \tilde{\alpha}_m(\boldsymbol{k}_m - \boldsymbol{q}) e^{-i c |\boldsymbol{k}_n| t} \right. \\
&\quad + |\boldsymbol{k}_n| |\boldsymbol{k}_m| \tilde{\alpha}_n(\boldsymbol{k}_n + \boldsymbol{q}) \tilde{\alpha}_m^*(\boldsymbol{k}_m + \boldsymbol{q}) e^{-i c |\boldsymbol{k}_n| t} \\
&\quad + |\boldsymbol{k}_n| |\boldsymbol{k}_m| \tilde{\alpha}_n^*(\boldsymbol{k}_n - \boldsymbol{q}) \tilde{\alpha}_m(\boldsymbol{k}_m - \boldsymbol{q}) e^{i c |\boldsymbol{k}_n| t} \\
&\quad - \left. |\boldsymbol{k}_n| |\boldsymbol{k}_m| \tilde{\alpha}_n^*(\boldsymbol{k}_n - \boldsymbol{q}) \tilde{\alpha}_m^*(\boldsymbol{k}_m + \boldsymbol{q}) e^{i c |\boldsymbol{k}_n| t} \right).
\end{aligned}
\end{equation}

The frequency domain current correlation function can be obtained by Fourier transform,
\begin{equation}
C(\boldsymbol{q}, \omega) = \int dt \, C(\boldsymbol{q}, t) e^{-i \omega t}.
\end{equation}
Substitute the expression for \( C(\boldsymbol{q}, t) \),
\begin{equation}
\begin{aligned}
C(\boldsymbol{q}, \omega) &= c^2 \sum_n \sum_m \left( - |\boldsymbol{k}_n| |\boldsymbol{k}_m| \tilde{\alpha}_n(\boldsymbol{k}_n + \boldsymbol{q}) \tilde{\alpha}_m(\boldsymbol{k}_m - \boldsymbol{q}) \int dt \, e^{-i (c |\boldsymbol{k}_n| + \omega) t} \right. \\
&\quad + |\boldsymbol{k}_n| |\boldsymbol{k}_m| \tilde{\alpha}_n(\boldsymbol{k}_n + \boldsymbol{q}) \tilde{\alpha}_m^*(\boldsymbol{k}_m + \boldsymbol{q}) \int dt \, e^{-i (c |\boldsymbol{k}_n| + \omega) t} \\
&\quad + |\boldsymbol{k}_n| |\boldsymbol{k}_m| \tilde{\alpha}_n^*(\boldsymbol{k}_n - \boldsymbol{q}) \tilde{\alpha}_m(\boldsymbol{k}_m - \boldsymbol{q}) \int dt \, e^{-i (\omega - c |\boldsymbol{k}_n|) t} \\
&\quad - \left. |\boldsymbol{k}_n| |\boldsymbol{k}_m| \tilde{\alpha}_n^*(\boldsymbol{k}_n - \boldsymbol{q}) \tilde{\alpha}_m^*(\boldsymbol{k}_m + \boldsymbol{q}) \int dt \, e^{-i (\omega - c |\boldsymbol{k}_n|) t} \right).
\end{aligned}
\end{equation}
Finally, evaluate the integrals using orthogonality of exponential functions,
\begin{equation}
\begin{aligned}
C(\boldsymbol{q}, \omega) &= c^2 \sum_n \sum_m \left( - |\boldsymbol{k}_n| |\boldsymbol{k}_m| \tilde{\alpha}_n(\boldsymbol{k}_n + \boldsymbol{q}) \tilde{\alpha}_m(\boldsymbol{k}_m - \boldsymbol{q}) 2\pi \delta(\omega + c |\boldsymbol{k}_n|) \right. \\
&\quad + |\boldsymbol{k}_n| |\boldsymbol{k}_m| \tilde{\alpha}_n(\boldsymbol{k}_n + \boldsymbol{q}) \tilde{\alpha}_m^*(\boldsymbol{k}_m + \boldsymbol{q}) 2\pi \delta(\omega + c |\boldsymbol{k}_n|) \\
&\quad + |\boldsymbol{k}_n| |\boldsymbol{k}_m| \tilde{\alpha}_n^*(\boldsymbol{k}_n - \boldsymbol{q}) \tilde{\alpha}_m(\boldsymbol{k}_m - \boldsymbol{q}) 2\pi \delta(\omega - c |\boldsymbol{k}_n|) \\
&\quad - \left. |\boldsymbol{k}_n| |\boldsymbol{k}_m| \tilde{\alpha}_n^*(\boldsymbol{k}_n - \boldsymbol{q}) \tilde{\alpha}_m^*(\boldsymbol{k}_m + \boldsymbol{q}) 2\pi \delta(\omega - c |\boldsymbol{k}_n|) \right).\label{ccffunction}
\end{aligned}
\end{equation}
The current correlation function above content four terms that corresponding to four momentum conserved processes. The first term describe annihilating two excitations at \(\boldsymbol{q}\) and \(-\boldsymbol{q}\), the second term describe firstly create and then annihilate an excitation at \(-\boldsymbol{q}\), the third and forth are conjugate processes of the first two, that create and then annihilate an excitation at \(\boldsymbol{q}\) and create two excitations at \(\boldsymbol{q}\) and \(-\boldsymbol{q}\).

\section{Fourier Transform of Elastic Constant Distribution Function} \label{sec. Fourier Transform of Elastic Constant Distribution Function}

The homogeneously distributed potential energy density requires zero gradient, that is,
\begin{equation}
    \nabla\langle \mathcal{E}_\text{pot}(\boldsymbol{r}) \rangle_t \equiv 0.
\end{equation}
Using the expression of Eq.\eqref{potdensityfunction} which is,
\begin{equation}
    \langle \mathcal{E}_\text{pot}(\boldsymbol{r}) \rangle_t = \pi G(\boldsymbol{r}) \sum_{\boldsymbol{k}_n} \left( \left( \nabla \alpha_n(\boldsymbol{r}) \right)^2 + |\boldsymbol{k}_n|^2\alpha_n(\boldsymbol{r})^2 \right).
\end{equation}
One can easily obtain the following relation between elastic constant \(G(\boldsymbol{r})\) and vibrational amplitude \( \alpha_n(\boldsymbol{r}) \) except a constant factor,
\begin{equation}
G(\boldsymbol{r})^{-1} = \sum_{\boldsymbol{k}_n} \left( |\boldsymbol{k}_n|^2 \alpha_n(\boldsymbol{r})^2 +  \nabla \alpha_n(\boldsymbol{r})^2 \right). \label{realspacerelation}
\end{equation}
It will be convenient to obtain the relation in Fourier space for use in the calculation of the current correlation function. The Fourier transform on the spatial domain of Eq.\eqref{realspacerelation} is given by,
\begin{equation}
\tilde{G}^{-1}(\boldsymbol{q}) = \int d^3r \, G(\boldsymbol{r})^{-1} e^{-i \boldsymbol{q} \cdot \boldsymbol{r}}.
\end{equation}
Substitute the expression for \( G(\boldsymbol{r})^{-1} \),
\begin{equation}
\tilde{G}^{-1}(\boldsymbol{q}) = \int d^3r \, \sum_{\boldsymbol{k}_n} \left( |\boldsymbol{k}_n|^2 \alpha_n(\boldsymbol{r})^2 + \left(\nabla \alpha_n(\boldsymbol{r})\right)^2 \right) e^{-i \boldsymbol{q} \cdot \boldsymbol{r}}.
\end{equation}
Separate the integrals,
\begin{equation}
\tilde{G}^{-1}(\boldsymbol{q}) = \sum_{\boldsymbol{k}_n} \left( |\boldsymbol{k}_n|^2 \int d^3r \, \alpha_n(\boldsymbol{r})^2 e^{-i \boldsymbol{q} \cdot \boldsymbol{r}} + \int d^3r \, \left(\nabla \alpha_n(\boldsymbol{r})\right)^2 e^{-i \boldsymbol{q} \cdot \boldsymbol{r}} \right).
\end{equation}
For the Fourier transform of \( \alpha_n(\boldsymbol{r})^2 \), let,
\begin{equation}
\tilde{\alpha}_n(\boldsymbol{q}) = \int d^3r \, \alpha_n(\boldsymbol{r}) e^{-i \boldsymbol{q} \cdot \boldsymbol{r}}.
\end{equation}
Using the convolution theorem, we have,
\begin{equation}
\int d^3r \, \alpha_n(\boldsymbol{r})^2 e^{-i \boldsymbol{q} \cdot \boldsymbol{r}} = \int \frac{d^3p}{(2\pi)^3} \tilde{\alpha}_n(\boldsymbol{p}) \tilde{\alpha}_n^*(\boldsymbol{q} - \boldsymbol{p}).
\end{equation}
For the Fourier transform of \( \left(\nabla \alpha_n(\boldsymbol{r})\right)^2 \), we use the property of the Fourier transform of the gradient,
\begin{equation}
\nabla \rightarrow i \boldsymbol{q},
\end{equation}
it becomes,
\begin{equation}
\int d^3r \, \left(\nabla \alpha_n(\boldsymbol{r})\right)^2 e^{-i \boldsymbol{q} \cdot \boldsymbol{r}} =- \int \frac{d^3p}{(2\pi)^3} \boldsymbol{p} \tilde{\alpha}_n(\boldsymbol{p}) (\boldsymbol{q}-\boldsymbol{p}) \tilde{\alpha}_n^*(\boldsymbol{q} - \boldsymbol{p}).
\end{equation}
Thus, the Fourier transform of the inverse elastic constant distribution function is,
\begin{equation}
\begin{aligned}
\tilde{G}^{-1}(\boldsymbol{q}) & = \sum_{\boldsymbol{k}_n} |\boldsymbol{k}_n|^2 \int \frac{d^3p}{(2\pi)^3} \tilde{\alpha}_n(\boldsymbol{p}) \tilde{\alpha}_n^*(\boldsymbol{q} - \boldsymbol{p}) - \int \frac{d^3p}{(2\pi)^3} \boldsymbol{p} \tilde{\alpha}_n(\boldsymbol{p}) (\boldsymbol{q}-\boldsymbol{p}) \tilde{\alpha}_n^*(\boldsymbol{q} - \boldsymbol{p})\\
&=\sum_{\boldsymbol{k}_n} |\boldsymbol{k}_n|^2 \tilde{\alpha}_n \star \tilde{\alpha}_n^* (\boldsymbol{q}) - [\boldsymbol{q} \tilde{\alpha}_n] \star [\boldsymbol{q} \tilde{\alpha}_n^*](\boldsymbol{q}), \label{relation}
\end{aligned}
\end{equation}
where \(\star\) denotes convolution. In later sections, this relation will be used for solving \(\tilde{\alpha}_n(\boldsymbol{q})\) from a given elastic constant distribution function.

\section{Solve the vibrational amplitude through a given elastic constant distribution function} \label{sec. Solve the vibrational amplitude through a given elastic constant distribution function}
We assume the distribution function of elastic constant inverse in wave vector space is a Delta function with a wave vector shift \(\boldsymbol{k}_G\),
\begin{equation}
    G^{-1}(\boldsymbol{q}) = \delta(\boldsymbol{k}_G + \boldsymbol{q}).
\end{equation}
When \(\boldsymbol{k}_G = 0\), \(G^{-1}(\boldsymbol{r})\) and hence \(G(\boldsymbol{r})\) beth are constant, that means the elastic constant have no fluctuations. When \(\boldsymbol{k}_G \neq 0\), the elastic constant will exhibit a fluctuation length \(1/|\boldsymbol{k}_G|\). In principle, \(G^{-1}(\boldsymbol{q})\) can have many components in wave vector space, \(G^{-1}(\boldsymbol{q}) =\sum_i p_i \delta(\boldsymbol{k}_{G,i} + \boldsymbol{q})\) with \(p_i\) a distribution probability. In the following, we will focus only on one component \(\boldsymbol{k}_G\) to avoid tedious repeating notations and enormous expansions. 

The following steps are trying to solve \(\tilde{\alpha}_n(\boldsymbol{q})\) from Eq.\eqref{relation},
\begin{equation}
\tilde{G}^{-1}(\boldsymbol{q}) = \sum_{\boldsymbol{k}_n} |\boldsymbol{k}_n|^2 \tilde{\alpha}_n \star \tilde{\alpha}_n^* (\boldsymbol{q}) - [\boldsymbol{q} \tilde{\alpha}_n] \star [\boldsymbol{q} \tilde{\alpha}_n^*](\boldsymbol{q}). 
\end{equation}
with the given \(G^{-1}(\boldsymbol{q}) = \delta(\boldsymbol{k}_G + \boldsymbol{q}) \). Noticing that the Delta function \(\delta(\boldsymbol{k}_G + \boldsymbol{q}) \neq 0\) only when \(\boldsymbol{q} = -\boldsymbol{k}_G\), and the convolution properties of Delta function,
\begin{equation}
\begin{aligned}
   &  [x \delta(x+a)] \star [x \delta(x+a)] = \int y \delta(y+a) (x-y) \delta(x-y+a)dy = -(ax + a^2)\delta(x+2a), \\
   & \delta(x+a) \star \delta(x+a) = \int \delta(y+a) \delta(x-y+a)dy = \delta(x+2a),
\end{aligned}
\end{equation}
where \(a\) is a constant. The above expression can be recast as,
\begin{equation}
\begin{aligned}
    \delta(\boldsymbol{k}_G + \boldsymbol{q}) = & \sum_{\boldsymbol{k}_n} \left( |\boldsymbol{k}_n|^2 - \dfrac{1}{2} \boldsymbol{k}_G \cdot \boldsymbol{q} - \dfrac{1}{4}\boldsymbol{k}_G^2 \right) \\
    & \times\dfrac{1}{N} \dfrac{1}{|\boldsymbol{k}_n|^2 - \dfrac{3}{4}\boldsymbol{k}_G^2} \delta\left(\dfrac{1}{2}\boldsymbol{k}_G + \boldsymbol{q}\right) \star \delta \left(\dfrac{1}{2}\boldsymbol{k}_G + \boldsymbol{q}\right). 
\end{aligned}
\end{equation}
Therefore one can solve the vibrational amplitude to be,
\begin{equation}
    \tilde{\alpha}_n(\boldsymbol{q}) = i\sqrt{\dfrac{1}{N}\dfrac{1}{|\boldsymbol{k}_n|^2 - \dfrac{3}{4}\boldsymbol{k}_G^2}} \delta(\dfrac{1}{2}\boldsymbol{k}_G + \boldsymbol{q}), \label{alphaq}
\end{equation}
with \(N\) is the normalization factor denoted the number of total modes. It is a pure imaginary function.

\section{Current correlation function for the given elastic constant distribution} \label{sec. Current correlation function for the given elastic constant distribution}
In this section, we compute the current correlation function of Eq.\eqref{ccffunction} with the vibrational amplitude given by Eq.\eqref{alphaq}. The current correlation function is,
\begin{equation}
\begin{aligned}
C(\boldsymbol{q}, \omega) &= c^2 \sum_n \sum_m |\boldsymbol{k}_n| |\boldsymbol{k}_m| \tilde{\alpha}_n^*(\boldsymbol{k}_n - \boldsymbol{q}) \tilde{\alpha}_m(\boldsymbol{k}_m - \boldsymbol{q}) 2\pi \delta(\omega - c |\boldsymbol{k}_n|) \\
&\quad - |\boldsymbol{k}_n| |\boldsymbol{k}_m| \tilde{\alpha}_n^*(\boldsymbol{k}_n - \boldsymbol{q}) \tilde{\alpha}_m^*(\boldsymbol{k}_m + \boldsymbol{q}) 2\pi \delta(\omega - c |\boldsymbol{k}_n|) + c.c.,
\end{aligned}
\end{equation}
where \(c.c.\) denotes the complex conjugate terms. Using the expression,
\begin{equation}
\tilde{\alpha}_n(\boldsymbol{q}) = i \sqrt{\dfrac{1}{N}\dfrac{1}{|\boldsymbol{k}_n|^2 - \dfrac{3}{4}\boldsymbol{k}_G^2}} \delta\left(\dfrac{1}{2}\boldsymbol{k}_G + \boldsymbol{q}\right),
\end{equation}
the current correlation function is,
\begin{equation}
\begin{aligned}
C(\boldsymbol{q}, \omega) &= c^2 \sum_n \sum_m |\boldsymbol{k}_n| |\boldsymbol{k}_m| (i) \sqrt{\dfrac{1}{N}\dfrac{1}{|\boldsymbol{k}_n|^2 - \dfrac{3}{4}\boldsymbol{k}_G^2}} \delta\left(\boldsymbol{k}_n - \left(\dfrac{1}{2}\boldsymbol{k}_G +\boldsymbol{q}\right)\right) \\
&\quad \times (-i) \sqrt{\dfrac{1}{N}\dfrac{1}{|\boldsymbol{k}_m|^2 - \dfrac{3}{4}\boldsymbol{k}_G^2}} \delta\left(\boldsymbol{k}_m - \left(\dfrac{1}{2}\boldsymbol{k}_G +\boldsymbol{q}\right)\right) 2\pi \delta(\omega - c |\boldsymbol{k}_n|) \\
&\quad - |\boldsymbol{k}_n| |\boldsymbol{k}_m| (-i) \sqrt{\dfrac{1}{N}\dfrac{1}{|\boldsymbol{k}_n|^2 - \dfrac{3}{4}\boldsymbol{k}_G^2}} \delta\left(\boldsymbol{k}_n - \left(\dfrac{1}{2}\boldsymbol{k}_G +\boldsymbol{q}\right)\right) \\
&\quad \times (-i) \sqrt{\dfrac{1}{N}\dfrac{1}{|\boldsymbol{k}_m|^2 - \dfrac{3}{4}\boldsymbol{k}_G^2}} \delta\left(\boldsymbol{k}_m + \left(\dfrac{1}{2}\boldsymbol{k}_G +\boldsymbol{q}\right)\right) 2\pi \delta(\omega - c |\boldsymbol{k}_n|) \\
& \quad + c.c..
\end{aligned}
\end{equation}
Simplify the expression,
\begin{equation}
\begin{aligned}
C(\boldsymbol{q}, \omega) &= \dfrac{c^2}{N} \sum_n \sum_m \dfrac{|\boldsymbol{k}_n| |\boldsymbol{k}_m|}{\sqrt{(|\boldsymbol{k}_n|^2 - \dfrac{3}{4}\boldsymbol{k}_G^2)(|\boldsymbol{k}_m|^2 - \dfrac{3}{4}\boldsymbol{k}_G^2)}} \\
&\quad \times \left[ \delta\left(\boldsymbol{k}_n - \left(\dfrac{1}{2}\boldsymbol{k}_G +\boldsymbol{q}\right)\right) \delta\left(\boldsymbol{k}_m - \left(\dfrac{1}{2}\boldsymbol{k}_G +\boldsymbol{q}\right)\right) 2\pi \delta(\omega - c |\boldsymbol{k}_n|) \right. \\
&\quad + \left. \delta\left(\boldsymbol{k}_n - \left(\dfrac{1}{2}\boldsymbol{k}_G +\boldsymbol{q}\right)\right) \delta\left(\boldsymbol{k}_m + \left(\dfrac{1}{2}\boldsymbol{k}_G +\boldsymbol{q}\right)\right) 2\pi \delta(\omega - c |\boldsymbol{k}_n|) \right] \\
& \quad + c.c.. \label{ccfgivenall}
\end{aligned}
\end{equation}

To this step, the current correlation function under fluctuating elastic constant with fluctuating length \(1/k_G\) have been obtained in Eq.\eqref{ccfgivenall}, which is one of the main conclusions of this work.

When \(\boldsymbol{k}_G = 0\), which indicates homogeneous elasticity, 
\begin{equation}
    \dfrac{|\boldsymbol{k}_n| |\boldsymbol{k}_m|}{\sqrt{(|\boldsymbol{k}_n|^2 - \dfrac{3}{4}\boldsymbol{k}_G^2)(|\boldsymbol{k}_m|^2 - \dfrac{3}{4}\boldsymbol{k}_G^2)}} = 1,
\end{equation}
Eq.\eqref{ccfgivenall} gives,
\begin{equation}
    C(\boldsymbol{q}, \omega) = \dfrac{2\pi c^2}{N}\delta(\omega - c|\boldsymbol{q}|).
\end{equation}
By doing the angular average and integral current correlation function over 3D wave vector space, the vibrational density of states (VDOS) can be obtained as,
\begin{equation}
\begin{aligned}
    D(\omega) & = \int d^3\boldsymbol{q} \, C(\boldsymbol{q}, \omega) \\
    & = \int dq \, C(q, \omega) 4\pi q^2 \\
    & = \int dq \, \dfrac{2\pi c^2}{N}\delta(\omega - cq) 4\pi q^2 \\
    & = \dfrac{8\pi^2}{N c} \omega^2,
\end{aligned}
\end{equation}
which recover the Debye's theory for the VDOS of homogeneous solids.

When \(\boldsymbol{k}_G \neq 0\), which indicates inhomogeneous elasticity with fluctuating length \(1/k_G\), direct computation of Eq.\eqref{ccfgivenall} is complicated due to involving \(\dfrac{1}{2} \boldsymbol{k}_G\) wave vector shift. However, if the wave vector shift is much smaller than Debye's wave vector, \(|\boldsymbol{k}_G| \ll k_D\), the wave vector shift in Delta function in Eq.\eqref{ccfgivenall} can be negligible even for large wave vector close to \(k_D\) and hence there is still \(|\boldsymbol{k}_n| \equiv |\boldsymbol{k}_m|\). Therefore the current correlation function becomes,
\begin{equation}
\begin{aligned}
    C(\boldsymbol{q}, \omega) & \approx \dfrac{2 \pi}{N} \dfrac{\omega^2}{\sqrt{(\omega^2 - c^2 \dfrac{3}{4}\boldsymbol{k}_G^2)(\omega^2 - c^2\dfrac{3}{4}\boldsymbol{k}_G^2)}}  \sum_n \delta\left(\boldsymbol{k}_n - \boldsymbol{q}\right) \\
    & \approx \dfrac{2 \pi}{N} \dfrac{\omega^2}{\omega^2 - c^2 \dfrac{3}{4}k_G^2} \sum_n  \delta\left(\boldsymbol{k}_n - \boldsymbol{q}\right).
\end{aligned}
\end{equation}
In the first step, the Delta function \(\delta(\omega - c|\boldsymbol{k}_n|)\) is calculated to replace \(|\boldsymbol{k}_n|\) in the factor. Doing the angular average and integral current correlation function over 3D wave vector space, the vibrational density of states (VDOS) can be obtained as,
\begin{equation}
\begin{aligned}
    D(\omega) & \approx \dfrac{2 \pi}{N} \int dq \, \dfrac{\omega^2}{\omega^2 + c^2 \dfrac{3}{4}k_G^2} 4\pi q^2 \\
    & \approx \dfrac{8\pi^2}{N c} \dfrac{\omega^4}{\omega^2 + c^2 \dfrac{3}{4}k_G^2}.
\end{aligned}
\end{equation}

Under low frequency \(\omega \ll c\sqrt{\dfrac{3}{4}}k_G\),
\begin{equation}
    \dfrac{\omega^4}{\omega^2 + c^2 \dfrac{3}{4}k_G^2} \rightarrow \omega^4,
\end{equation}
under high frequency \(\omega \gg c\sqrt{\dfrac{3}{4}}k_G\),
\begin{equation}
    \dfrac{\omega^4}{\omega^2 + c^2 \dfrac{3}{4}k_G^2} \rightarrow \omega^2.
\end{equation}
The frequency \(c\sqrt{\dfrac{3}{4}}k_G\) is where power law change, and hence the frequency of Boson peak.

Another way to to see the effect of nonzero \(|\boldsymbol{k}_G|\) is to limit there is only one allowed mode in the displacement intensity field \(\phi(\boldsymbol{r}, t)\), that means,
\begin{equation}
\phi(\boldsymbol{r}, t) = \alpha_0(\boldsymbol{r}) e^{-i (\boldsymbol{k}_0 \cdot \boldsymbol{r} - c |\boldsymbol{k}_0| t)} + \alpha_0^*(\boldsymbol{r}) e^{i (\boldsymbol{k}_0 \cdot \boldsymbol{r} - c |\boldsymbol{k}_0| t)}.
\end{equation}
Then using Eq.\eqref{ccfgivenall} the current correlation function becomes,
\begin{equation}
    C(\boldsymbol{q}, \omega) = \dfrac{2\pi c^2}{N} \dfrac{k_0^2}{k_0^2 + \dfrac{3}{4}k_G^2} \delta\left(\boldsymbol{k}_0 - \left(\dfrac{1}{2}\boldsymbol{k}_G + \boldsymbol{q}\right)\right) \delta(\omega - ck_0).
\end{equation}
Therefore the \(|\boldsymbol{k}_G|\) dressed vibrational mode will appear only at,
\begin{equation}
    C\left(\boldsymbol{k}_0 - \dfrac{1}{2}\boldsymbol{k}_G, ck_0 \right) = \dfrac{2\pi c^2}{N} \dfrac{k_0^2}{k_0^2 + \dfrac{3}{4}k_G^2},
\end{equation}
instead of standard acoustic phonon mode \(\left(\boldsymbol{k}_0, ck_0 \right)\). This wave vector shift may correspond to the quasi-flat dispersion relation observed recently in experiments and simulations.

\end{document}